\documentclass[%
 reprint,
 unsortedaddress,
 nobibnotes,
 amsmath,amssymb,
 aps,
 superscriptaddress,
 floatfix,
]{revtex4-2}

\usepackage{graphicx}
\usepackage{bm}
\usepackage{xfrac}
\usepackage{mathptmx} 
\usepackage{soul}

\newcommand{\wt}{\widetilde}
\newcommand{\ave}[1]{{\left<#1\right>}}
\newcommand{\inv}[1]{\frac{1}{#1}}
\DeclareSymbolFont{newfont}{OML}{cmm}{m}{it} 
\DeclareMathSymbol{\Varrho}{3}{newfont}{37}
\newcommand{\abs}[1]{{\left|#1\right|}}
\newcommand{\rmd}{\text{d}}
\newcommand{\dtau}{\Delta}
\newcommand{\tmin}{\tau_\downarrow}
\newcommand{\tmax}{\tau_\uparrow}
\newcommand{\taudim}{\ensuremath{s}}
\newcommand{\taud}{\langle \taudim \rangle}
\newcommand{\taunond}{\tau}
\newcommand{\normalizFac}{\eta}
\newcommand{\Phiave}{\ensuremath{\ave{\Phi}}}
\newcommand{\Phirms}{\ensuremath{\Phi}_\text{rms}}
\newcommand{\Phiwt}{\ensuremath{\widetilde{\Phi}}}
\newcommand{\betaEst}{\bar{\beta}}

\newcommand{\Eqref}[1]{Eq.~\eqref{#1}}
\newcommand{\Eqsref}[1]{Eqs.~\eqref{#1}}
\newcommand{\Figref}[1]{Fig.~\ref{#1}}
\newcommand{\Figsref}[1]{Figs.~\ref{#1}}

\begin{document}

\preprint{APS/123-QED}

\title{Apparent universality of $1/f$ spectra as an artifact of finite-size effects}

\author{M.~A.~Korzeniowska}
 \email{magdalena.a.korzeniowska@uit.no}
\author{A.~Theodorsen}
 \email{audun.theodorsen@uit.no}
\affiliation{Department of Physics and Technology, UiT The Arctic University of Norway, N-9037 Troms{\o}, Norway.}
\author{M.~Rypdal}
 \email{martin.rypdal@uit.no}
\affiliation{Department of Mathematics and Statistics, UiT The Arctic University of Norway, N-9037 Troms{\o}, Norway.}
\author{O.~E.~Garcia}
 \email{odd.erik.garcia@uit.no}
\affiliation{Department of Physics and Technology, UiT The Arctic University of Norway, N-9037 Troms{\o}, Norway.}

\date{\today}

\begin{abstract}
Power spectral density scaling with frequency $f$ as $1/f^\beta$ and $\beta \approx 1$ is widely found in natural and socio-economic systems. Consequently, it has been suggested that such self-similar spectra reflect the universal dynamics of complex phenomena. Here, we show that for a superposition of uncorrelated pulses with a power-law distribution of duration times the estimated scaling exponents $\betaEst$ depend on the system size. We derive a parametrized, closed-form expression for the power spectral density, and demonstrate that for $\beta \in [0,2]$ the estimated scaling exponents have a bias towards $\betaEst=1$. For $\beta=0$ and $\beta=2$ the explicit logarithmic corrections to frequency scaling are derived. The bias is particularly strong when the scale invariance spans less than four decades in frequency. Since this is the case for the majority of empirical data, the boundedness of systems well described by the superposition of uncorrelated pulses may contribute to overemphasizing the universality of $1/f$.
\end{abstract}

\maketitle

\textit{Introduction.--- }%
A wide range of complex systems display spatial or temporal scale invariance, fractality, and long-range dependence (LRD) 
\cite{1985-Dennis-SoPh..100..465D, 1999-Boffetta-PhysRevLett.83.4662, 2003-Sanchez-PhysRevLett.90.185005, 2006-Arcangelis-PhysRevLett.96.051102, 2021-Aschwanden-AstrophJ, 2005-Paczuski-PhysRevLett.95.181102, 2018-Tindale-JGeophRes, 1983-Pellegrini-PhysRevB.27.1233, 2013-Liu-ApplPhysLett, 2019-Tadic, 2020-Franzke-RevGeoph, 2016-Rypdal-esd-7-281-2016, 2006-Huybers-Nature, 1983-Mandelbrot-book, 1997-Bak-book, 1991-Schroeder-book}. In particular, the emergence of self-similar frequency power spectral density scaling $1/f^\beta$ has been of interest since the discovery of a $1/f$-type noise in vacuum tubes almost a century ago \cite{1926-Schottky-PhysRev.28.74, 1925-Johnson-PhysRev.26.71}. Reports of scaling exponents $\beta$ close to unity in various systems have led to questions about universality. Theoretical ideas such as self-organized criticality (SOC) have been put forward \cite{1987-Bak-PhysRevLett.59.381}. However, identifying a general mechanism for the observed variety of self-similar behavior has proved difficult \cite{1999-DeLosRios-PhysRevLett.82.472, 2014-Chamberlin-PhysRevE.90.012142, 2017-Yadav-PhysRevE.96.022215, 2010-Eliazar-PhysRevE.82.021109, 2019-De-PhysRevB.99.024305, 2009-Nardone-PhysRevB.79.165206}.

In this paper, we demonstrate that an apparent $1/f$ universality arises in a generalized filtered Poisson process subject to finite-size effects \cite{2018-Loscar-PhysRevE.97.032103, 2013-Niemann-PhysRevLett.110.140603}. The shot-noise approach is canonical for the phenomenological modeling of LRD statistics of fluctuating systems, from background noise to violent bursts \cite{1988-Bak-PhysRevA.38.364, 1989-Jensen-PhysRevB.40.7425, 2005-Lowen-book, 2011-Aschwanden-book, 2016-Samorodnitsky-book, 2017-Pipiras-book}. We derive a closed-form expression for the parametrized power spectral density of a finite-size system and explore its scale invariance while varying the self-similarity range and the exponent $\beta\in[0,2]$. We assess the finite-size effects by comparing the asymptotic scaling relations with the effective scaling of the analytical power spectral density. Our results show that the observed scaling is always biased towards $\beta=1$ in the presence of finite-size effects, and the bias is most substantial when the scaling range is narrow.

\textit{Filtered Poisson process.--- }%
Let us first introduce the theoretical framework for our analysis. Consider a stochastic process given by a superposition of $K$ uncorrelated, independent and identically distributed pulses $\phi(\theta)$, occurring as a random sequence in a time interval of duration $T$ \cite{2017-garcia-PhysPlasm}, 
\begin{equation}\label{PhiK_shotnoise}
    \Phi_K(t) = \sum_{k=1}^{K(T)} A_k\phi\left( \frac{t-t_k}{\taudim_k} \right) .
\end{equation}
Each pulse labeled $k$ is characterized by an amplitude $A_k$, a duration time $\taudim_k$, and an arrival time $t_k$ distributed uniformly on the interval $T$. The pulse-duration times are assumed to be randomly distributed with probability density $P_\taudim(\taudim)$, and an average pulse-duration time $\langle \taudim \rangle = \int_0^\infty \rmd\taudim\,\taudim\,P_\taudim(\taudim)$. Given the distribution of pulse amplitudes $P_A(A)$, we use Campbell's theorem to compute the moments and the autocorrelation function of the process \eqref{PhiK_shotnoise} by averaging over all random variables for the case of exactly $K$ pulses \cite{1909-Campbell-ProcCambr, 2017-garcia-PhysPlasm}, and subsequently averaging over the randomly distributed number of pulses $K$. This yields the rigorous characteristics of the stationary process $\Phi(t)$ \cite{1972-Butz-JStatPhys}. The power spectral density follows directly as the Fourier transform of the autocorrelation function. For the standardized process $\Phiwt = (\Phi-\Phiave)/\Phirms$, and with a normalized, dimensionless duration time $\taunond = \taudim/\taud$, the power spectral density is expressed in a non-dimensional form as
\begin{equation}\label{psdphifullydimless}
    \Omega_{\wt{\Phi}}(\omega) = \int_0^\infty \rmd\tau\,\tau^2 P_{\tau}(\tau) \Varrho_\phi(\tau\omega),
\end{equation}
where $\omega=2\pi f \taud$ denotes the dimensionless angular frequency, $\Varrho_\phi(\taunond\omega)=\int_{-\infty}^\infty \rmd\theta\,\rho_\phi(\theta)\exp(-i\taunond\omega\theta)$ is the Fourier transform of the normalized autocorrelation function $\rho_\phi$ of the pulse function $\phi$, and $P_{\taunond}(\taunond) = \taud\,P_\taudim(\taudim)$ is the normalized probability density function for pulse durations \cite{2017-garcia-PhysPlasm}.

\textit{Pareto distributed durations.--- }%
Equation \eqref{psdphifullydimless} holds for an arbitrary finite-mean distribution $P_\taunond(\taunond)$ of pulse durations. In particular, it holds for a bounded Pareto distribution with exponent $\alpha$ and a finite support $\left[\tmin, \tmax\right]$, normalized by a factor $\normalizFac(\tmin,\tmax,\alpha)$ such that $\int_0^\infty \rmd\taunond\,P_\taunond(\taunond)=1$,
\begin{equation} \label{bpareto}
    P_\taunond(\taunond;\tmin,\tmax,\alpha) =
    \begin{cases}
        \displaystyle \normalizFac\,\taunond^{-\alpha}  & \text{if }\tmin \leq \taunond \leq \tmax , \\
    	0 & \text{otherwise}.
    \end{cases}
\end{equation}
The normalization of $P_\taunond$ and the inherent property of a normalized-variable mean $\langle \taunond \rangle = \int_{\tmin}^{\tmax} \rmd\taunond\,\taunond P_\taunond(\taunond) = 1$ put two constraints on the three parameters $\{\tmin$, $\tmax$, $\alpha$\} in \Eqref{bpareto}. Defining a dimensionless ratio parameter $\dtau = \tmax/\tmin$ and solving the resulting system of three constraints, we obtain $\tmin$, $\tmax$ and $\eta$ in terms of $\alpha$ and $\dtau$ as
\begin{subequations}\label{eq:tau_eta}
\begin{align}
    \tmin(\dtau,\alpha) & = \tfrac{(\alpha-2) (1 - \dtau^{1 - \alpha})}{(\alpha-1) (1 - \dtau^{2 - \alpha})} ,
    \\
    \tmax(\dtau,\alpha) & = \dtau\tmin ,
    \\
    \normalizFac(\dtau,\alpha) & = \tfrac{(\alpha-1)}{1 - \dtau^{1-\alpha}} \tmin^{\alpha-1}, \label{tmax_eta}
\end{align}
\end{subequations}
with well-defined limits for $\alpha\rightarrow1$ and $\alpha\rightarrow2$. Given \Eqsref{eq:tau_eta}, the probability distribution given by \Eqref{bpareto} is parametrized as $P_\taunond = P_\taunond(\taunond;\dtau,\alpha)$.

We note that a finite, nondivergent mean $\ave{\taunond}=1$ is a requirement for the stationarity of the process given by \Eqref{PhiK_shotnoise}, and the well-defined normalization of the power spectral density given by \Eqref{psdphifullydimless}. 
With the chosen parametrization $P_\taunond(\taunond;\dtau,\alpha)$ and the condition $\ave{\taunond}=1$, the effect of the increase in $\dtau$ on the boundaries $\tmin$ and $\tmax$ depends on the value of $\alpha$.
When $\alpha < 1$ the divergence $\dtau\rightarrow\infty$ is driven by the decrease $\tmin\rightarrow0$, rather than by the increase of $\tmax$, thus hindering long-range correlations. 
As $\alpha\rightarrow0$, $P_\taunond(\taunond)$ given by \Eqref{bpareto} reduces to a uniform distribution, with finite mean and variance \cite{2017-garcia-PhysPlasm}.

\textit{Scale invariance.--- }%
In the unbounded limit, $P_\taunond(\taunond)$ defined by \Eqref{bpareto} displays self-similar scaling
\begin{equation}
    \lim_{\substack{\tmin\rightarrow0 \\ \tmax\rightarrow\infty}} P_\taunond(\lambda\taunond) = \lim_{\substack{\tmin\rightarrow0 \\ \tmax\rightarrow\infty}} \lambda^{-\alpha}P_\taunond(\taunond),\label{Ptau_scaling}
\end{equation}
which together with \Eqref{psdphifullydimless} implies a power-law scaling relation for the power spectral density,
\begin{equation}\label{psdscaling}
    \lim_{\substack{\tmin\rightarrow0 \\ \tmax\rightarrow\infty}} \Omega_{\wt{\Phi}}(\lambda\omega) = \lim_{\substack{\tmin\rightarrow0 \\ \tmax\rightarrow\infty}} \lambda^{\alpha-3}\,\Omega_{\wt{\Phi}}(\omega).
\end{equation}
Equation \eqref{psdscaling} suggests the existence of a universal $1/\omega^\beta$ self-similarity of the power spectral density given by \Eqref{psdphifullydimless}, with $\beta(\alpha)=3-\alpha$. Strictly, the probability distribution given by \Eqref{bpareto} is not well defined in the asymptotic limit, but bounding $\tmin$ at an arbitrarily small value results in a finite variance of the process for $\alpha>3$, and an infinite variance otherwise. In order to ensure a finite pulse-duration mean in the asymptotic limit, $\alpha\geq1$ is required. Thus, we conjecture that if $\Omega_{\wt{\Phi}}$ displays a power-law signature in the limit when $\dtau\rightarrow\infty$, then it does so for Pareto exponents $1\leq\alpha\leq3$. The resulting power spectral density scaling exponents range within $0\leq\beta(\alpha)\leq2$. Exponents $\alpha=1$, $\alpha=2$, and $\alpha=3$ characterize Brownian, pink, and white noise signatures with $\beta=2$, $\beta=1$, and $\beta=0$, respectively. 

The spectral scale invariance of a finite-size system is confined to the frequency range limited by the cutoff values $\omega\tmax=1$ and $\omega\tmin=1$, ranging over $\log_{10}\dtau$ decades in frequency. Outside this range the power spectral density assumes the shape determined by the power spectra of the pulse function $\phi$, following a broken power law with the associated break points to and from the $1/\omega^\beta$ scaling.

\begin{figure*}
    \includegraphics{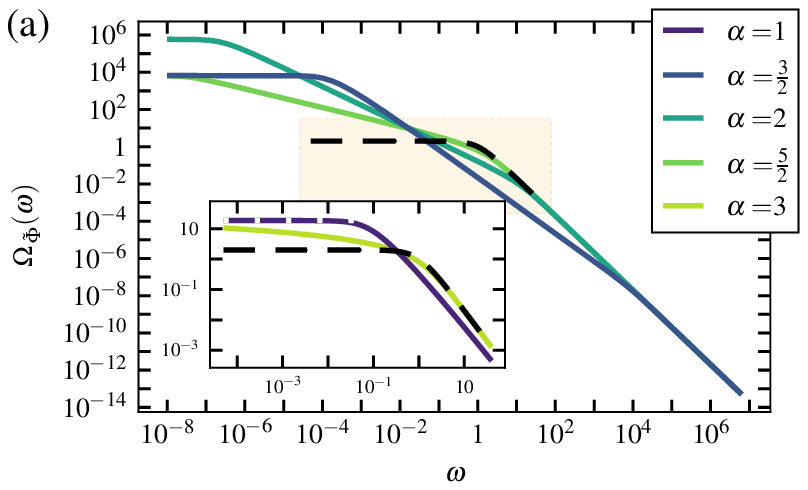}\qquad
    \includegraphics{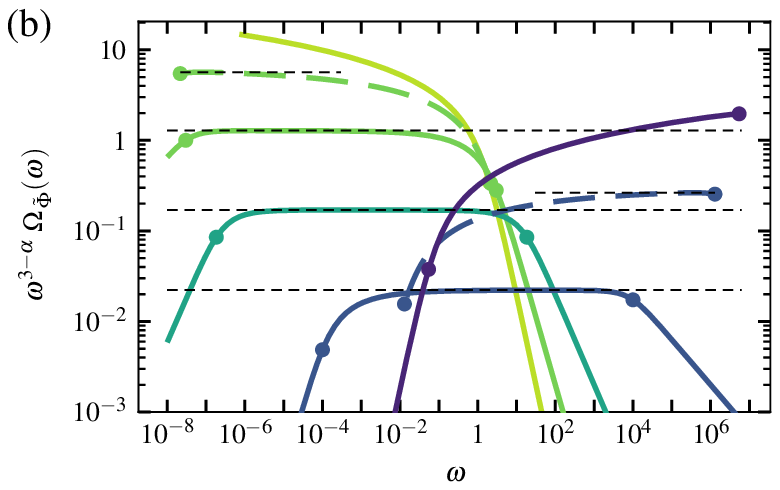}
    \\
    \includegraphics{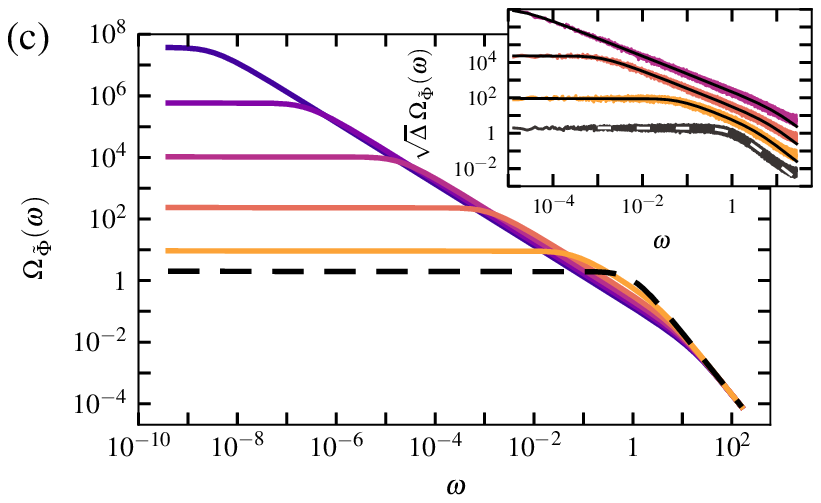}\qquad
    \includegraphics{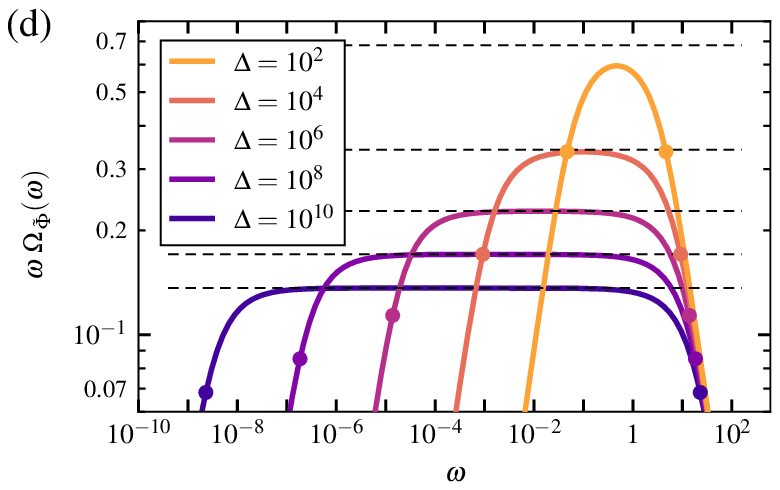}
    \caption{
        Frequency power spectral density of the filtered Poisson process with one-sided exponential pulse shape and Pareto-distributed pulse-duration times. Legend color coding applies per row. 
        \textbf{{Top row}}: Varied $\alpha$ at fixed $\dtau=10^8$. 
        \textbf{{Bottom row}}: Varied $\dtau$ at fixed $\alpha=2$.
        \textbf{{Left column}}: Uncompensated spectra $\Omega_{\wt{\Phi}}(\omega;\dtau, \alpha)$ given by \Eqref{psdphi-explicit}. Dashed lines represent Lorentzian-function spectra. 
        \textbf{{Right column}}: Compensated spectra $\omega^{3-\alpha}\,\Omega_{\wt{\Phi}}(\omega;\dtau, \alpha)$. 
        The horizontal dashed black lines spanning the entire $\omega$ range mark the inverse of the compensating prefactors according to \Eqsref{psdlim}. The regions where the dashed black lines overlap with the colored lines indicate the ranges of power-law scaling. 
        Colored dots mark the theoretical boundaries of the self-similarity ranges, $\omega\tmax=1$ and $\omega\tmin=1$. 
        \textbf{(a)} The inset presents the spectra at the boundaries of the LRD regime, $\alpha=1$ $(\beta=2)$ and $\alpha=3$ $(\beta=0)$, where logarithmic corrections to $1/\omega^\beta$ scaling apply. The domain represented in the inset is shaded in the outer plot.
        \textbf{(b)} Two ancillary $\alpha$ cases plotted with dashed colored lines showcase the reduction in the range of self-similarity when $\alpha$ is $1/7$ away from the nearest LRD boundary.
        \textbf{(c)} The inset presents the empirical power spectra obtained for realizations of the process, shifted vertically by a factor $\sqrt{\dtau}$ to avoid overlapping. The color coding of the empirical spectra is aligned to the legend. 
        The overlying solid black lines represent the corresponding analytical results.
        An additional empirical case $\dtau=0$, representing a constant pulse duration, is plotted in black and overlaid by a dashed-white Lorentzian.
        }
    \label{fig:wide-alpha-delta-scans}
\end{figure*}

\textit{Power-law spectra.--- }%
The asymptotic scaling relation $\beta=3-\alpha$ is verified for a one-sided exponential pulse function $\phi$, 
\begin{equation}\label{phi-exp}
    \phi(\theta) = 
    \begin{cases}
        \exp(-\theta) & \text{if } \theta \geq 0 ,\\
        0 & \text{otherwise},
    \end{cases}
\end{equation}
whose power spectral density follows to be a Lorentzian function $\Varrho_\phi(\vartheta) = 2/(1+\vartheta^2)$ \cite{2017-garcia-PhysPlasm}. For a constant pulse duration $\taunond=\ave{\taunond}$ the power spectral density given by \Eqref{psdphifullydimless} inherits the Lorentzian shape $\Omega_{\wt{\Phi}}(\omega) = 2\ave{\taunond}/(1+\ave{\taunond}^2\omega^2)$, flat for low frequencies and with a $1/\omega^2$ tail for high frequencies, consistent with $\beta\rightarrow0$ and $\beta\rightarrow2$, respectively. For distributed pulse durations, \Eqsref{psdphifullydimless}, \eqref{bpareto} and \eqref{eq:tau_eta}
yield an explicit, closed-form expression for the frequency power spectral density parametrized by $\dtau$ and $\alpha$: 
\begin{multline} \label{psdphi-explicit}
    \Omega_{\wt{\Phi}}(\omega;\dtau,\alpha) = \\
    \begin{cases}
        \frac{1}{\ln\dtau\,\omega^2}
        \ln\left(\frac{(\dtau-1)^2 + \dtau^2 \ln^2\dtau\, \omega^2}{(\dtau-1)^2 + \ln^2\dtau\, \omega^2}\right) & \text{if }\alpha=1, \\
        \frac{2} {\ln\dtau\,\omega }
        \left[\arctan\left(\frac{(\dtau - 1)\omega}{\ln\dtau}\right) 
        - \arctan\left(\frac{(\dtau - 1)\omega} {\dtau \ln\dtau}\right)\right] & \text{if } \alpha = 2, \\
        \frac{2}{(\dtau^\alpha-\dtau)\omega^2}\,\left[ 
        \dtau^\alpha\;_2F_1\left( 1,\frac{\alpha-1}{2},\frac{\alpha+1}{2};-\inv{\tmin^2\omega^2} \right) \right.  \\ 
        \qquad\qquad\quad \left. - \dtau\;_2F_1\left(1,\frac{\alpha-1}{2},\frac{\alpha+1}{2};-\inv{\tmax^2\omega^2} \right) \right] & \text{otherwise},
    \end{cases}\\ \raisetag{0pt}
\end{multline}
where $_2F_1$ is a hypergeometric function defined by Gauss series \cite{hyp2f1}. The expected frequency scaling $1/\omega^{3-\alpha}$ is manifested by considering the compensated spectra in the limit of an infinitely broad distribution of duration times. For several values of $\alpha$ representing the LRD regime $1\leq\alpha\leq3$, the following \Eqsref{psdlim} present both the prefactors and the powers of $\omega$ which together satisfy the compensation of the power spectral density $\Omega_{\wt{\Phi}}(\omega;\dtau, \alpha)$ given by \Eqref{psdphi-explicit}, 
\begin{subequations}\label{psdlim}
\begin{align}
    \lim_{\dtau\rightarrow\infty} \Omega_{\wt{\Phi}}(\omega;\dtau,1) 
    &\quad \;\; \frac{\ln{\dtau}}{\ln\left(\omega^2\ln^2{\dtau} \right)}
    & &\;\omega^2 
    &= 1 ,\label{psdlim-alpha1}\\
    \lim_{\dtau\rightarrow\infty} \Omega_{\wt{\Phi}}(\omega;\dtau, \tfrac{3}{2})
    &\quad \;\; \frac{\sqrt{2}(\sqrt{\dtau}-1)}{\pi\sqrt[4]{\dtau}}
    & &\abs{\omega}^{3/2}
    &= 1 ,\label{psdlim-alpha1-5}\\
    \lim_{\dtau\rightarrow\infty} \Omega_{\wt{\Phi}}(\omega;\dtau,2)
    &\quad \qquad\; \frac{\ln{\dtau}}{\pi}
    & &\abs{\omega}
    &= 1 ,\label{psdlim-alpha2}\\
    \lim_{\dtau\rightarrow\infty} \Omega_{\wt{\Phi}}(\omega;\dtau,\tfrac{5}{2})
    &\quad \frac{\sqrt{6} (\sqrt{\dtau}-1)}{\pi \sqrt{1 + \sqrt{\dtau} + \dtau}} 
    & &\abs{\omega}^{1/2} 
    &= 1 ,\label{psdlim-alpha2-5}\\
    \lim_{\dtau\rightarrow\infty} \Omega_{\wt{\Phi}}(\omega;\dtau,3)
    &\quad  2 \left[\ln{\left(1+\tfrac{4}{\omega^2}\right)}\right]^{-1} 
    & & &= 1 \label{psdlim-alpha3}.
\end{align}
\end{subequations}
Equation \eqref{psdlim-alpha2} reveals the $1/\omega$ signature of the pink noise, obtained for $\alpha=2$. Logarithmic corrections to the theoretical frequency scaling are present at the LRD-regime boundaries, $\alpha=1$ and $\alpha=3$. Similar logarithmic corrections have been linked to phase transitions and critical behavior of certain statistical-mechanical systems \cite{2006-Kenna-PhysRevLett.96.115701, 2010-Sandvik-PhysRevLett.104.177201, 2020-Hong-PhysRevE.101.012124}, as well as demonstrated for a renewal process with power-law-distributed waiting times \cite{1993-Lowen-PhysRevE.47.992}.

The parameters $\alpha$ and $\dtau$ represent two mechanisms shaping the power spectral density in the range of self-similarity: logarithmic corrections and boundedness. Figures \ref{fig:wide-alpha-delta-scans}(a) and \ref{fig:wide-alpha-delta-scans}(c) present plots of the power spectral density $\Omega_{\wt{\Phi}}(\omega;\dtau,\alpha)$ given by \Eqref{psdphi-explicit} for multiple choices of $\alpha$ and $\dtau$, respectively. The corresponding compensated spectra are presented in \Figsref{fig:wide-alpha-delta-scans}(b) and \ref{fig:wide-alpha-delta-scans}(d). The chosen values of $\alpha$ span the entire LRD regime, and are aligned to \Eqsref{psdlim}. The selected values of $\dtau$ allow for examining the scaling behavior of $\Omega_{\wt{\Phi}}(\omega;\dtau,\alpha)$ over different ranges of self-similarity.
Compensated spectra aid the identification of the power-law scaling.

\textit{Logarithmic corrections.--- }%
Figure \ref{fig:wide-alpha-delta-scans}(b) confirms the existence of power-law scaling for $\alpha=\sfrac{3}{2}$, $\alpha=2$, and $\alpha=\sfrac{5}{2}$, 
as well as the logarithmic corrections to scaling at the boundaries of the LRD regime, $\alpha=1$ and $\alpha=3$. The curvature of the compensated spectra increases as $\alpha$ moves away from the center of the LRD regime, $\alpha=2$, causing gradual shortening of the power-law scaling ranges. The dashed colored lines in \Figref{fig:wide-alpha-delta-scans}(b) reveal the shape of the compensated spectra for $\alpha=2\pm\sfrac{6}{7}$ ($\beta=1\mp\sfrac{6}{7}$), equivalent to $1/7$ away from the nearest LRD-regime boundary. These two cases demonstrate that the loss of power-law scaling occurs already inside the LRD regime, not only at its boundaries.

\textit{Boundedness.--- }%
The theoretical boundaries of the power-law scaling ranges, given by \Eqref{eq:tau_eta}, are marked with dots in \Figsref{fig:wide-alpha-delta-scans}(b) and \ref{fig:wide-alpha-delta-scans}(d). The broken power laws affect the spectral scaling in the vicinity of $\omega\tmax=1$ and $\omega\tmin=1$ by reducing the effective ranges of self-similarity. Figure \ref{fig:wide-alpha-delta-scans}(d) shows that in the center of the LRD regime, $\alpha=2$, the reduction is by approximately one and a half frequency decades on each side of the self-similarity range, for any of the considered values of $\dtau$. Power-law scaling does not emerge unless the underlying process is characterized by at least four decades ($\dtau\geq10^4$) of scale invariance.

The empirical power spectral densities obtained for realizations of the stochastic process given by \Eqref{PhiK_shotnoise} expectedly match the corresponding analytical predictions given by \Eqref{psdphi-explicit}. Examples for $\alpha=2$ and different values of $\dtau$ are shown in the inset in \Figref{fig:wide-alpha-delta-scans}(c).

\begin{figure}
    \includegraphics{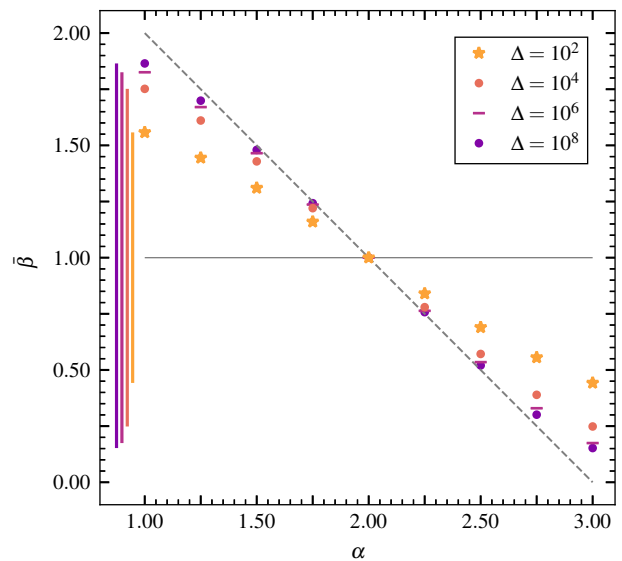}
    \caption{
        Estimated power-law scaling exponents $\betaEst$ of the analytical power spectral density curves $\Omega_{\wt{\Phi}}(\omega;\dtau,\alpha)$ given by \Eqref{psdphi-explicit} for various ranges $\dtau$ of the underlying scale invariance, and in the entire LRD regime $1\leq\alpha\leq3$.
        The dashed gray line marks the asymptotic scaling relation $\lim_{\dtau\rightarrow\infty}\beta(\alpha) = 3-\alpha$. 
        The solid gray line marks $\betaEst=1$ representative of the $1/f$ noise.
        The colorful vertical sidebars mark the range of $\betaEst$ observed for different values of $\dtau$.
        Legend color coding is aligned to \Figref{fig:wide-alpha-delta-scans}(d).
        }
    \label{fig:Beta--3-alpha}
\end{figure}

\textit{Apparent universality.--- }%
The combined effect of the logarithmic corrections to frequency scaling and the boundedness of the self-similarity range is gauged by comparing the effective scaling of the analytical power spectral density $\Omega_{\wt{\Phi}}(\omega;\dtau, \alpha)$ given by \Eqref{psdphi-explicit} for various combinations of the parameters $\alpha$ and $\dtau$, to the asymptotic scaling relation $\lim_{\dtau\rightarrow\infty}\beta(\alpha) = 3-\alpha$. In order to reduce the effect of the break-point curvature, half a decade is discarded on each side of the theoretical self-similarity range, shifting the boundaries of the power-law fitting ranges to $\omega\tmax=10^{1/2}$ and $\omega\tmin=10^{-1/2}$, respectively. Linear least-square fits are made to logarithmically spaced points in double-logarithmic coordinates. The resulting estimations of power-law scaling exponents $\betaEst$ are presented in \Figref{fig:Beta--3-alpha}. As $\alpha$ approaches any of the LRD-regime boundaries, the effective $\betaEst(\alpha)$ relation diverges from the asymptotic limit $\beta(\alpha)=3-\alpha$ towards the central value $\betaEst=1$. The divergence is stronger for small $\dtau$.

The colored sidebars in \Figref{fig:Beta--3-alpha} mark the ranges of the estimated exponents $\betaEst$ for different values of $\dtau$.
For $\dtau=10^8$ the range is $\betaEst\approx1\pm0.86$. We recall that \Figref{fig:wide-alpha-delta-scans}(b) demonstrates a notable curvature of the compensated spectra for $\dtau=10^8$ and $\alpha=2\pm\sfrac{6}{7}$ ($\beta=1\mp0.86$). For $\dtau=10^2$ and $\dtau=10^4$ we further recall that even at the center of the LRD regime, $\alpha=2$ ($\beta=1$), the compensated spectra in \Figref{fig:wide-alpha-delta-scans}(d) reveal none, or very short power-law scaling ranges, respectively. The lack of power-law scaling does not affect the power-law fitting procedure. The estimated exponents range within $\betaEst\approx1\pm0.56$ for $\dtau=10^2$, and $\betaEst\approx1\pm0.75$ for $\dtau=10^4$.

The findings presented in \Figsref{fig:wide-alpha-delta-scans} and \ref{fig:Beta--3-alpha} indicate that the effective spectral scaling is biased towards $\betaEst=1$, and the bias increases with the decrease of $\dtau$, or with $\alpha$ approaching the LRD-regime boundaries. Specifically:
(1) For the ranges of the underlying scale invariance shorter than approximately four decades ($\dtau<10^4$) the power spectral density does not display power-law scaling.
(2) For the longer $\dtau$ ranges the spectral power-law scaling is manifested only for a subrange of exponents centered around $\alpha=2$ ($\beta=1$).
(3) The extent of this sub-range increases with the increase of $\dtau$, up to the asymptotic limit $\alpha\in(1,3)$ [$\beta\in(0,2)$] when $\dtau\rightarrow\infty$.

\textit{Discussion.--- }%
The results presented in \Figref{fig:Beta--3-alpha} are obtained under favorable conditions: Power-law fitting is made to logarithmically spaced data points following analytical curves, exact boundaries of the self-similarity ranges are known, and symmetric cutoffs are applied to reduce the effect of the break-point curvature. Despite these measures the effective $\betaEst(\alpha)$ relation is biased towards $\betaEst=1$ with respect to the asymptotic $\lim_{\dtau\rightarrow\infty}\beta(\alpha)=3-\alpha$. The scaling exponents close to the LRD-regime boundaries $\beta=0$ and $\beta=2$ are not observed for any of the investigated finite values of $\dtau$.

The power spectral density of a one-sided exponential pulse has asymptotic scaling as $1/\omega^0$ for low frequencies and $1/\omega^{2}$ for high frequencies. The associated break points in the spectrum affect the self-similarity range, biasing the underlying $1/\omega^\beta$ scaling towards $\betaEst=1$. The wider the range for power-law fitting, the more weight is put on the break-point curvature. 
Experiments show that discarding significant margins on both sides of the fitting range reduces the bias, yielding more accurate scaling estimations when compared with the theoretical predictions. However, for relatively narrow ranges of scale invariance the break-point curvature affects the entire $1/\omega^\beta$ range, inflicting a bias too extensive to retrieve the underlying $1/\omega^\beta$ scaling. 
Consulting compensated spectra allows for scrutinizing the effective scale invariance.

Narrow ranges of scale invariance prone to the $\betaEst\rightarrow1$ bias may overemphasize the universality of $1/f$-type scaling. Observing long ranges of scale invariance demands both that the underlying process is long-range self-similar, and that it is measured with precision and scope satisfying the long-range extent \cite{2013-Niemann-PhysRevLett.110.140603}. Estimating power-law statistics of unequally sampled or merged data sets has been addressed in Refs. \cite{2014-Lovejoy-ClimDyn, 2019-Navas-Portella-PhysRevE}.

If the exact boundaries of the self-similarity range are not known, the choice of the power-law fitting range is arbitrary, and possibly biased towards either low or high frequencies. Different methods of spectral scaling estimation may increase the bias, or compensate for it. The smoothness of the effective $\betaEst(\alpha)$ relations presented in \Figref{fig:Beta--3-alpha} suggests that knowing the boundaries of the self-similarity range might facilitate tracing back from the observed scaling to the underlying scaling of the studied process.

\textit{Conclusions.--- }%
The results presented here demonstrate that the estimated spectral scaling of long-range dependent processes may be biased towards $1/f$ in the presence of finite-size effects. This bias results from the curvature in the spectra due to broken power-law scaling, as well as the logarithmic corrections associated with long range dependence. Identification of the true power-law scaling requires scale invariance over several decades in frequency in the underlying process, as shown in \Figref{fig:wide-alpha-delta-scans}(d). Empirical data seldom display accordingly broad ranges of self-similarity \cite{2018-Tindale-JGeophRes, 1983-Pellegrini-PhysRevB.27.1233, 2013-Liu-ApplPhysLett, 2019-Tadic, 2020-Franzke-RevGeoph, 2016-Rypdal-esd-7-281-2016}, suggesting a spectral scaling bias at least in the case of processes that are well described by a superposition of uncorrelated pulses. Considering that a variety of physical phenomena has been canonically modeled in this way \cite{1988-Bak-PhysRevA.38.364, 1989-Jensen-PhysRevB.40.7425, 2005-Lowen-book, 2011-Aschwanden-book, 2016-Samorodnitsky-book, 2017-Pipiras-book}, the observed $1/f$ universality may be overstated. Whether a similar bias is present for other complex-dynamics systems requires further investigation.

\begin{acknowledgments}
    This work was supported by the UiT Aurora Centre Program, UiT The Arctic University of Norway (2020). A.~T.~was supported by Troms{\o} Research Foundation under Grant No. 19\_SG\_AT.
\end{acknowledgments}

%


\end{document}